# Defects in the β-$Ga_2O_3$($\bar{2}$01)/$HfO_2$ MOS system and the effect of thermal treatments

*Khushabu. S. Agrawal[1], Paolo LaTorraca[1], Jonas Valentijn[1], Roberta Hawkins[2,‡], Adam A. Gruszecki[2], Joy Roy[2], Vasily Lebedev[3], Lewys Jones[3], Robert M. Wallace[2], Chadwin D. Young[2], Paul K. Hurley[1] and Karim Cherkaoui[1,\*]*

[1]Tyndall National Institute, University College Cork, Lee Maltings, Prospect Row, Cork, Ireland

[2]Department of Materials Science and Engineering, The University of Texas at Dallas, Richardson, Texas, 75080, United States

[3]Advanced Microscopy Laboratory (AML), School of Physics, CRANN & AMBER Trinity College Dublin, the University of Dublin, Dublin 2, Ireland

[‡] Now at Qorvo, Inc., Richardson, Texas, 75080, United States

ABSTRACT: **We have investigated the properties of the β-$Ga_2O_3$($\bar{2}$01)/$HfO_2$/Cr/Au MOS (metal-oxide-semiconductor) system after annealing (450°C) in different ambient conditions (forming gas, $N_2$ and $O_2$). Defect properties have been analyzed using an approach combining experimental impedance measurements with physics-based simulations of the capacitance-voltage (C-V) and conductance-voltage (G-V) characteristics of β-$Ga_2O_3$/$HfO_2$ MOS capacitors. This approach enabled us to detect two defect bands in $HfO_2$ characterized by thermal ionization energies of ~1.1eV (acceptor-like) and ~2eV (donor-like) attributed to a polaronic self-trapping state and an oxygen vacancy in $HfO_2$, respectively. This study demonstrates how thermal treatments affect the energy distributions and densities of the observed defects. The adopted methodology also enabled the extraction of the spatial distribution of defects across the $HfO_2$ thickness and Cr/$HfO_2$ interface. The high concentration of oxygen vacancies close to the Cr/$HfO_2$ interface extracted from experimental and simulated electrical data is confirmed by *in-situ* XPS analysis which shows how Cr is scavenging oxygen from the $HfO_2$ and creating the donor band confined near the Cr/$HfO_2$ interface. This donor band density is observed to be reduced after annealing as per simulation and unchanged for different annealing conditions. We speculate this may be due to the formation of dense films and polyforms of $HfO_2$ under different ambient as revealed by high-resolution TEM images.**

KEYWORDS: *β-$Ga_2O_3$, Ultra-wide band gap, Hafnium oxide, Defects*

1. Introduction

β-Ga$_2$O$_3$ has been widely explored for high power applications because of its superior electrical properties compared to SiC and GaN, including a wider band gap of ~4.8 eV and a high breakdown field that is 2-3 times greater (~8 MVcm$^{-1}$) [1,2]. In addition, the cost-effective production of β-Ga$_2$O$_3$ is another significant advantage of β-Ga$_2$O$_3$-based high power device applications. The recent trends in β-Ga$_2$O$_3$-based metal-oxide-semiconductor devices explore the use of HfO$_2$, Al$_2$O$_3$ and SiO$_2$ dielectrics [3–6]. Realization of Ga$_2$O$_3$ MOS devices faces some severe challenges such as charge trapping or the absence of hole conduction which restricts the design to unipolar devices or heterojunctions.

High dielectric constant HfO$_2$ material has been introduced in the fabrication of Si-based complementary MOS technology, which has enhanced device performance and lowered power consumption. HfO$_2$ possesses a high dielectric constant (k~16-25) [7,8], relatively high conduction band offsets and higher barrier heights with respect to gate metals. High-k dielectrics are also desirable for electric field management in lateral power devices [9–11]; therefore, it is noteworthy to consider HfO$_2$ as a dielectric material for Ga$_2$O$_3$-based MOS devices. SiO$_2$ and Al$_2$O$_3$ have lower k values than HfO$_2$ but offer higher band offsets and higher band gaps resulting in lower leakage current and better thermal gate stack stability [12–14].

The properties of the dielectric/semiconductor and dielectric/metal interfaces have a great impact on the overall electrical properties of the device. Dielectric oxide defects may also significantly alter the device characteristics and reliability. Interface traps, and border traps are the most common defects in high-k dielectrics which can affect shorter/longer term performance degradation of the device. Oxygen vacancies in HfO$_2$ lead to the threshold voltage shift and even lower the carrier mobility in the channel material[15]. Defects in $Al_2O_3$ are responsible for Fermi level pinning and a substantial threshold voltage shift in capacitance-voltage (C-V) [16], whereas the interface state density between $SiO_2/Ga_2O_3$ causes the C-V to stretch out and introduces frequency dispersion in accumulation capacitance[17].

These defects result from the dielectric deposition process or during treatments after dielectric deposition such as rapid thermal annealing (RTA)[18], which influences the C-V characteristics. The impact of high temperature process annealing on $HfO_2/\beta$-$Ga_2O_3$ has been investigated by Masten et al., and their findings showed how high temperature annealing at 700-850°C increases the leakage current and density of interface traps [19]. Jayawardhana et al. demonstrate how annealing of $Ga_2O_3/Al_2O_3$ MOS devices at 500°C for 2 min in forming gas (FG) causes a rise in shallow interface states in comparison with the annealing in $N_2$ ambient[20]. Commonly reported problems include C-V flat band shifting following consecutive sweeps, C-V stretching, and dispersion in accumulation capacitance where the origin remains unspecified after annealing. Thus, to progress technology for high power applications using $Ga_2O_3$-based MOS devices, it is important to investigate the source of these limitations as annealing is an essential step in the fabrication process.

In this work, we have systematically detailed how defects in the $HfO_2/\beta$-$Ga_2O_3$ system play a significant role in changing the C-V properties after annealing in various ambient conditions in comparison with the as-deposited (control sample). A thorough analysis has been conducted, which details how annealing affects the space-energy distribution of traps within hafnia and primarily alters the MOS device's C-V and G-V characteristics. We replicate the experimental data using the Ginestra simulation platform, analyzing the distribution of defects in the oxide and interface with $Ga_2O_3$ and Cr metal. The extracted thermal ionization energy of the two defect bands indicates the presence of an intrinsic polaronic defect near the $Ga_2O_3/HfO_2$ interface and the oxygen vacancy in $HfO_2$ near the $HfO_2/Cr$ interface. [21,22]. We also investigated the $HfO_2/Cr$ interface with the help of *in-situ* X-ray photoelectron spectroscopy (XPS) explaining Cr is scavenging oxygen from the $HfO_2$ layer.

## 2. Experimental Methods

The MOS structures were fabricated using unintentionally doped ($\bar{2}01$) $\beta$-$Ga_2O_3$ semiconductor substrates procured from Novel Crystal Technology, Japan. More information regarding ALD of $HfO_2$ and wafer pre-cleaning can be found elsewhere[23]. Following the ~12 nm $HfO_2$ ALD process, the substrate was split into 4 samples: Three samples were subjected to an RTA (post deposition annealing (PDA)) treatment at 450°C for 5 min in forming Gas (5% $H_2$ + 95% $N_2$), Oxygen ($O_2$) and Nitrogen ($N_2$) ambient, respectively, and one sample was used as a control sample (as-deposited). For all samples, circular contacts with 200 μm diameter of Cr (20 nm)/Au (150 nm) were formed as top gate electrodes through shadow mask stencils using E-Beam evaporation. 20 nm Ti and 150 nm Au layers were deposited as back contacts. $BCl_3$ etching was carried out prior to Ti/Au metal deposition. The schematic of the $\beta$-$Ga_2O_3/HfO_2/Cr$ MOS structure is shown in Figure 1 along with the MOS capacitor process flow. The C-V and G-V measurements were carried out using an Agilent E4980 LCR meter. Each C-V measurement was done at room temperature in a dark environment at frequencies ranging from 100 Hz to 1 MHz.

The *in-situ* XPS characterization was performed by forming the $Ga_2O_3/$ $HfO_2$(12nm)/Cr (1nm) samples in the ultra-high vacuum (UHV) cluster tool. The XPS in the cluster tool is equipped with a monochromatic Al Kα source and an Omicron EA125 hemispherical analyzer, achieving a resolution of ±0.05 eV. In this work, XPS was performed in as-loaded conditions and compared with those after subsequent process steps. A reference Cr layer (25 nm) was also deposited under UHV condition on a $SiO_2$/Si substrate to compare with the Cr reaction products. The details of the sample preparation are provided in the supplementary information. All the XPS spectra were fitted using Analyzer software[24]. The lithography process was avoided throughout the experiment to maintain clean, in-situ environments.

The high-resolution transmission electron microscopy (HRTEM) and energy dispersive analysis (EDS) were performed after characterizing the samples electrically. The lamellas were prepared by TESCAN Solaris Ga+ dual-beam focused ion beam (FIB) system. To protect the region of interest before FIB milling multiple protecting layers (50–100 nm carbon layer, followed by a 300–500 nm platinum layer and additional 3.5μm thick carbon protecting layer for protection) were deposited using electron beam induced deposition. Thinning was performed at 30 kV, with a final low-energy polish at 5 kV (20 pA) to minimize ion beam-induced damage. High-resolution TEM and STEM were performed using FEI Titan 80-300 microscope (Themofisher Scientific, USA) equipped with a Schottky field-emission gun operated at 300 kV, and a HAADF detector for STEM. TEM images were recorded with a Gatan UltraScan CCD camera (Gatan, Inc., USA) and processed using the open source Gwyddion software. EDS mappings were acquired using a Bruker XFlash 6-30 EDS detector (Bruker Co., USA), and processed using the open source Hyperspy library.

Simulations of the MOS capacitor impedance characteristics were performed using Applied Materials Ginestra simulation platform to extract the defect properties[25]. The software employs the calculation of charge transport, electrostatics in MOS governed by the direct tunneling, and trap assisted tunneling self consistently to/from the charge trapping/emission dynamics of the defects. More importantly, the charge transport phenomenon is modeled according to the multi-phonon trap assisted tunneling theory which explains the electron-phonon coupling and atomic lattice relaxation associated with the charge trapping systematically. Table I lists the general parameters used to reproduce the experimental results. The doping concentration in $Ga_2O_3$ was extracted from the $1/C^2$ versus V curve of the as-deposited sample.

Table I: Materials parameters used for the C-V and G-V simulations[2,26,27]

| | | |
|---|---|---|
| β-$Ga_2O_3$ | Band Gap (eV) | 4.85 |
| | Electron Affinity χ (eV) | 4 |
| | Doping $N_D$ ($cm^{-3}$) | $1.8 \times 10^{17}$ |
| | Relative Dielectric Permittivity ε | 10 |
| | Effective Electron Mass $m^*$ | $0.28 m_o$ |
| | Carrier mobility $\mu_n$ ($cm^2\ V^{-1}\ s^{-1}$) | 300 |
| Cr | Work Function (eV) | 4.5 |
| $HfO_2$ | Band Gap (eV) | 5.8 |
| | Electron affinity χ (eV) | 2.4 |
| | Effective electron tunnelling Mass $m_T$ | $0.25 m_o$ |
| | Oxide Thickness | 12 nm |

3. **Results and Discussion**

The flat band shift in $Ga_2O_3$ MOS C-V characteristics is often present irrespective of the dielectric used. The charge trapping in the $Al_2O_3/(\bar{2}01)Ga_2O_3$ MOS capacitors showed C-V stretch out and flat band shift after successive voltage sweep capacitance measurements[20]. Chen et al. have shown a similar stretch out behavior with $ZrO_2$ dielectric on $Ga_2O_3$ which was attributed to interface traps[28]. Further, Zeng et al. studied the interface states behavior in $SiO_2/Ga_2O_3$ MOS capacitors and relate this stretch out and flat band shift with defect states in the dielectric and at the interface[17]. In the current study, we have observed a significant flat band shift after successive C-V sweeps (not shown for brevity), which has also been observed by other groups[19,29–31]. This flat band shift, which we attribute to trapped charge upon successive voltage sweeps, is still present several days after the initial measurement (the systematic study of the trapped charge dynamics over time is beyond the scope of this paper). The origin of this behavior hasn't been extensively discussed in literature, and it is important to examine some of the causes of this behavior in the high-k/$Ga_2O_3$ system. We believe that this *semi-permanent* charge trapping effect is not only caused by the presence of defect states located deep in the dielectric presenting high thermal activation energy, but this effect is strongly dependent on the electric field distribution across the high-k/$Ga_2O_3$ device. In MOS capacitors including conventional semiconductors (energy band gap <2 eV), the electric field across the dielectric for a given bias in the inversion region is comparable with the field when the device is in the accumulation region. For $Ga_2O_3$ MOS devices the situation is different as the semiconductor does not invert due to its wide band gap of 4.85 eV, which gives rise to asymmetry in the electric field. The applied bias in depletion drops across the dielectric and $Ga_2O_3$ which results in strong electric field asymmetry between the accumulation and depletion conditions of the MOS system. This situation is compounded by the higher k value of $HfO_2$ (16) than the k value of $Ga_2O_3$ (10). To investigate this, we have simulated the electric fields in the inversion regime of $HfO_2$ on Si and $HfO_2$ on $Ga_2O_3$ MOS capacitors by considering zero defects in $HfO_2$ (ideal case). Figure 2 compares the electric field across both systems at specified bias of -2V which is in depletion (or inversion) mode of the MOS device. A Cr metal gate was assumed for the simulation, and the n-type doping level in Si and $Ga_2O_3$ ($2\times10^{17}$ $cm^{-3}$) and semiconductor affinity values were very similar (4 eV), which allows the comparison of the electric field across both structures for a given applied bias. At -2V in depletion/inversion, the electric field in $HfO_2$ is approximately 5 times lower for the $Ga_2O_3$ MOS structure in comparison with the $HfO_2$ field for the Si device. This situation implies that, for high-k/$Ga_2O_3$ devices to achieve dielectric field magnitudes in depletion comparable to dielectric fields in accumulation, a significantly higher bias in depletion is required. The implication for charge trapping in high-k/$Ga_2O_3$ MOS devices is that as the device is swept into accumulation, the trapped charge is not necessarily fully recovered when the device is biased in depletion regardless of the traps' concentration or their thermal ionization and often results in a flat band shift of the C-V with each successive voltage sweep measurement.

Figure 3 (a-d) shows the experimental C-V and G-V data curves for all 4 samples namely as-deposited, forming gas, $N_2$ and $O_2$ ambient annealed, respectively, measured at room temperature. These are the initial C-V curves measured at different frequencies (100 Hz to 1 MHz) under frequency sweep mode. In frequency sweep mode, measurement frequency is swept for every gate voltage. The bias is incremented by 0.1 V at each step and then capacitance and conductance values are taken at different frequencies. This will ensure that all frequency measurements are recorded in a single bias sweep and that the multi-frequency characteristics receive an identical bias stress. The as-deposited C-V curve (figure 3.a) shows significant frequency dispersion in accumulation. This dispersion is related to the capture and emission process of majority carriers by oxide traps located close to the oxide/semiconductor interface, while the parallel flat band shift towards positive voltage is associated with the negative oxide charge (fixed or trapped charge). As shown in Figure. 3(b), an increase in frequency dispersion and C-V stretch out after PDA in forming gas ambient for 5 min were observed. The relative flat band shift and decrease in accumulation capacitance after the PDA is also noted. Further stretch out in C-V, increase in frequency dispersion, and positive flat band voltage shifting with a decrease in accumulation capacitance were overserved for $N_2$ and $O_2$ annealed samples as shown in Figure.3 (c) and (d), respectively. This reveals the degradation of the $HfO_2$ properties and the generation of more defects in the oxide due to the PDA. The respective conductance voltage (G-V) characteristics for all four samples are also shown in Figure 3(e-h). The G-V characteristics show decrease in conductance values in deep depletion regime for annealed samples as compared to the control sample.

To extract the defect properties in $HfO_2$ for all four samples, we used Ginsetra which implements a multi-phonon trap assisted tunneling (MPTAT) model[32]. In this framework, the active defects are characterized by a capture cross section σ, phonon energy $\hbar\omega$, Huang-Rhys factor S, tunneling mass $m_T$ and trap levels localized in energy and space $E_{T,Z}$.[33] The relaxation energy $E_{REL}$ determines the capture emission time constant and temperature dependence which is useful to identify the physical nature of the defects. The parameters utilized for simulation are listed in Table I. It is well established that the oxygen vacancies are common bulk defects in $HfO_2$ causing the degradation of its dielectric properties and distort electrical characteristics in MOS devices[18,34,35]. The quantitative energy-space distributions of traps causing the degradation in experimental C-V were simulated by considering different traps taking part in the degradation process. We have considered two defect bands to match the experimental C-V and G-V characteristics: an acceptor-like band likely distributed across the full $HfO_2$ thickness, and a donor-like band spatially localized near the Cr electrode as shown in Fig. (4 a). Both acceptor-like (0/-1) and donor-like (+1/0) traps were considered for simulating the C-V, G-V and reproducing the experimental data in figure 3.

The defect energy distributions were extracted for all samples and plotted in figure 4b with the main defect parameters summarized in table II. Figure. 3 also compares the experimental C-V characteristics (symbols) with the simulated C-V (solid lines) for all four samples by considering the defects' energy distribution approximately 1.1 eV below the HfO$_2$ conduction band (Ec). Another defect (donor-like) band about 2 eV below the conduction band of HfO$_2$ is required to produce the best match with experimental G-V for the gate bias in the deep depletion region as shown in Figure 3(e-h). It is important to note that the acceptor band required to explain the multi-frequency C-V behavior in accumulation does not necessarily need to extend across the full HfO$_2$ thickness. At the bias, temperature, and frequency measurement conditions, electrons are only able to communicate with defects located up to approximately 5 nm from the HfO$_2$/Ga$_2$O$_3$ interface. However, the intrinsic acceptor-like traps is likely to be distributed across the HfO$_2$ and this assumption has no effect on the simulated C-V and G-V characteristics. The donor band, however, is found not to be uniformly spatially distributed. The donor-like traps are present up to 3 nm from the Cr interface; this constraint is imposed by the C-V characteristic, which isn't matched with a uniform donor distribution.

The defect density (acceptor and donor) varies between $5\times10^{19}$ to $5\times10^{20}$ cm$^{-3}$ for all the four samples while an E$_{REL}$ value of 1-1.2 eV was used for the simulation. To match the experimental data with the simulation different *k* values were required. It is worth noting that a k value of 17 gives the best match for the as-deposited sample, while *k* values in the range of 13-16 give the best match for forming gas, N$_2$ and O$_2$ samples. Such low-*k* values for HfO$_2$ are unexpected and the possible origin will be discussed in the next section.

Figure. 4(b) summarizes the effect of annealing conditions on the traps' distribution over energy and space within the band gap of HfO$_2$. It is observed that the donors' thermal ionization energy mean remains constant before and after thermal treatment while their density strongly decreases. The acceptors' density, however, increased significantly post thermal annealing associated with a slight increase in thermal ionization energy spread. The two effects combined result in more defect states aligned with the $Ga_2O_3$ conduction band and therefore much more distorted C-V after annealing.

**Table II: Hafnia Defects Parameters**

| Sample | Dielectric constant($k$) | Acceptor defect | Donor defect |
|---|---|---|---|
| Phonon energy ($E_{PH}$) | - | 0.03 eV | 0.06 eV |
| Capture cross-section ($\sigma$) | - | $5 \times 10^{-17}$ cm$^2$ | $10^{-14}$ cm$^2$ |
| As deposited | 17 | $E_{th}$: 1.1 eV<br>$E_{rel}$: 1eV<br>Std dev=0.2eV<br>Defect density: $6 \times 10^{19}$ cm$^{-3}$ | $E_{th}$: 2eV<br>$E_{rel}$: 1.2 eV<br>Std dev= 0.1 eV<br>Defect density: $5 \times 10^{20}$ cm$^{-3}$ |
| Forming Gas annealed | 16 | $E_{th}$: 1.1eV<br>$E_{rel}$: 1eV<br>Std dev=0.25eV<br>Defect density: $2 \times 10^{20}$ cm$^{-3}$ | $E_{th}$: 2eV<br>$E_{rel}$: 1.2 eV<br>Std dev=0.1 eV<br>Defect density: $5 \times 10^{19}$ cm$^{-3}$ |
| N$_2$ annealed | 14 | $E_{th}$: 1.1 eV<br>$E_{rel}$: 1 eV<br>Std dev=0.25eV<br>Defect density: $4 \times 10^{20}$ cm$^{-3}$ | $E_{th}$: 2eV<br>$E_{rel}$: 1.2 eV<br>Std dev= 0.1 eV<br>Defect density: $5 \times 10^{19}$ cm$^{-3}$ |
| O$_2$ annealed | 13 | $E_{th}$: 1.17eV<br>$E_{rel}$: 1.12eV<br>Std dev=0.2eV<br>Defect density: $3.5 \times 10^{20}$ cm$^{-3}$ | $E_{th}$: 2eV<br>$E_{rel}$: 1.2 eV<br>Std dev= 0.1 eV<br>Defect density: $5 \times 10^{19}$ cm$^{-3}$ |

Many studies on the application of HfO$_2$ dielectric in MOS devices confirm that one of the most prevalent defects currently observed is oxygen vacancies. These oxygen vacancies play an important role in generating trapping sites in MOS devices[15,22]. Annealing significantly influences these defects by modifying the material properties and inducing a phase transition in HfO$_2$ [36,37]. In addition, the concentration of these oxygen vacancies seems to be phase dependent. The amorphous HfO$_2$ tends to have significant oxygen vacancies due to a less ordered structure, while the phase transition to cubic may have more oxygen vacancies tolerance[38,39]. Another source of trapping states in both amorphous and crystalline $HfO_2$ are polaronic states, originating from the strong interaction between charge carriers and the lattice vibrations (i.e phonons) in the material. The oxygen vacancy may exist in +2, +1, 0, -1 and -2 states. The +2, +1 states are deep trap states and while the neutral state is expected to occur positioned ~ 1 eV below the CB minima[21]. These defects are common in $HfO_2$ regardless of the substrate on which it is deposited.

The donor defect band distributed near the Cr interface could be explained by the Cr strong oxygen affinity when deposited on $HfO_2$ or other oxides and could lead to oxygen scavenging from $HfO_2$ and increase the $V_O$ density locally[40,41]. *In-situ* XPS analysis was performed on $Cr/HfO_2/Ga_2O_3$ stack to provide more in-depth chemical information about the $Cr/HfO_2$ interface.

Figure 5 shows the XPS results before and after Cr deposition on $HfO_2$ dielectric grown on β-$Ga_2O_3$ ($\bar{2}01$) substrate (without annealing: control sample). Since the substrate signal was below the detection limit of XPS (see the supplementary file Figure S1: survey spectra) only the interface between Cr and $HfO_2$ is analyzed. Peak positions are referenced to the C 1s (284.8 eV) peak to minimize charging effects observed after oxide growth. Figure 5 (a) shows the Hf 4f and O 1s core level peaks after ALD deposition and following UHV metallization. Signals detected in both Hf 4f at 17.3 eV and O 1s at 530.4 eV binding energy (BE) denote $HfO_2$ state [42,43]. The sub stoichiometric state is below the detection limit of XPS for the $HfO_2$ surface. C and O 1s show surface organic impurities likely sourced from the ALD chamber where metal-organic precursors are used. C1s in Figure 5(c) demonstrates a small amount of carbon compounds at the surface consistent with O 1s [44]. No carbide species (~282 eV) [45] was detected following ALD deposition.

However, after Cr deposition multiple reactions were observed at the Cr/HfO$_2$ interface compared to reference Cr. (see Figure 5 (b) "Cr (Reference)"). In Figure 5(a), the full width at half maximum (FWHM) of Hf 4f increased by 0.12 eV after 1 nm Cr deposition and a shoulder peak near the lower BE site appeared. Similar changes are observed in O 1s suggesting a reduction of HfO$_2$ by Cr and the formation of a sub-stoichiometric (HfO$_2$-x) state. Cheng et al. reported a similar sub-stoichiometric state in the HfO$_2$/GaN interface[46]. Although thermodynamically HfO$_2$ is more stable than chromium oxides (approximate standard Gibbs free energy, $\Delta G°_f$, HfO$_2$ 548.5 kJ/mol per bond of Hf-O and $\Delta G°_f$, Cr$_2$O$_3$ =-379.7 kJ/mol per bond of Cr-O)[47], amorphous HfO$_2$ has a less stable atomic structure with higher defect densities[48]. This can alternatively make oxygen more mobile and easier to scavenge by Cr. Moreover, metal-induced defect states can be also responsible for such an oxygen scavenging effect[49]. Cr 2p in Figure 5(b) shows multiple oxidation states compared to the reference Cr metal. Along with metallic Cr$^0$ states, oxides and carbides [50]are also detected in Cr 2p. O 1s (530.6 eV) and C1s (~283.3 eV) show the Cr oxide and carbide [51]respectively suggesting the presence of aggressive reactions at the HfO$_2$/Cr interface.

A shift towards higher binding energy values in Hf 4f and O1s spectra was observed after metal deposition. The scavenging of oxygen from HfO$_2$ creates oxygen vacancies or related donor states which may eventually shift the oxide's fermi level (E$_F$) towards the conduction band [15,52]. Additionally, dipole formation at the Cr/ HfO$_2$ interface might also induce some BE shift[53]. The Cr fermi level alignment could also induce downward HfO$_2$ band bending and contribute to the BE shift.

Considering the effect of thermal annealing on the defect bands, the donor concentration is reduced by one order of magnitude after annealing but is unaffected by the different annealing conditions (as depicted in figure 4(b)). It seems the annealed HfO$_2$ films react less with Cr as compared to the as-deposited films (*In-situ* XPS analysis on annealed HfO$_2$ was not available for this study). On the other hand, the acceptor defect density is observed to increase for the annealed samples as compared to the control sample. Furthermore, N$_2$ and O$_2$ annealing conditions generate a greater number of acceptor defects than the FG annealed sample. The extracted trap distribution 1.1 eV below Ec supports the attribution of the acceptor to the polaronic defect in HfO$_2$[18,54]. These polaronic traps are associated with under-coordinated Hf ions or elongated Hf-O bonds. The deeper state polarons in polycrystalline films have the elongated Hf-O bonds confined to the disordered region[38].

The *k* value is observed to be dependent over the annealing conditions (17 for as deposited and 13 for O$_2$ annealed). Cross-sectional TEM images of the as deposited and annealed films (Figure. 6) confirm the thickness of HfO$_2$ layer is ~ 11.5 ± 0.5 nm for all four samples and show no evidence of interlayer formation between the Ga$_2$O$_3$ and HfO$_2$ and rule out HfO$_2$ thickness variation after annealing. The low HfO$_2$ *k* value extracted cannot be attributed to higher than nominal HfO$_2$ physical thickness. It is interesting to note that both as-deposited and annealed samples seem to be polycrystalline in nature. The fast Fourier transform patterns were analyzed to assess the value of the observed interplanar distances of the HfO$_2$ films. An observed d-spacing of 0.27 nm is in a good agreement with the literature data for crystalline HfO$_2$ (Figure S2) [55,56]. The extracted *k* value of 17 for the as-deposited sample is consistent with the monoclinic phase reported by several groups for HfO$_2$ on Si[57,58]; however, the cause for the lower *k* value (13-16) of the other samples is unclear. Ga diffusion into HfO$_2$ was observed previously in HfO$_2$/Ga$_2$O$_3$ samples annealed at high temperature[19]; however, EDS analysis doesn't reveal any significant Ga diffusion into HfO$_2$ in our samples. A possible source of the *k* value degradation could be the partial oxidation of the Cr layer in contact with HfO$_2$ (Figure S3 and S4).

## 4. Conclusion

We have characterized the defects properties in the $HfO_2/(\bar{2}01)\beta\text{-}Ga_2O_3$ system. The approach we have adopted, which combines the experimental electrical characteristics with physics-based simulation of MOS devices, was successful in profiling active defects in $Au/Cr/HfO_2/Ga_2O_3$ MOS devices. Two distinct defect bands with thermal ionization energies ~ 1.1 eV and ~2eV below the $HfO_2$ conduction band have been extracted. The acceptor band at $E_{th}=1.1$ eV has been associated to polaronic intrinsic defects in $HfO_2$ in good agreement with the earlier reports, its density and thermal ionization energy spread are observed to be increasing after annealing in different conditions, while the donor band with $E_{th}=2$ eV has been associated with the oxygen vacancy donor level in $HfO_2$ which is reduced after annealing and unchanged in different annealing conditions. This defect band is mainly localized near the $HfO_2$/Cr interface indicating that Cr may be at the origin of this defect by reducing the $HfO_2$ and generating oxygen vacancies, this hypothesis was supported by *in-situ* XPS analysis. The study shows how thermal treatment degrades the properties of $HfO_2/Ga_2O_3$ MOS capacitors and provides useful insights for the selection of process parameters and materials for future $Ga_2O_3$ based devices.

**Supporting Information**.

XPS sample preparation and Survey spectra scan (Figure S1); TEM and EDS characterization: Cross sectional TEM images for all 4 (control and annealed) samples showing the polycrystalline $HfO_2$(Figure S2). Compositional elemental mapping of different layers with EDS intensity normalized spectra (Figure S3 and S4). Cumulative EDS spectra (Figure S5.)

AUTHOR INFORMATION

**Corresponding Author**

* Karim Cherkoui

Email: karim.cherkaoui@tyndall.ie

**Author Contributions**

The manuscript was written through contributions of all authors. All authors have given approval to the final version of the manuscript.

**Funding Sources**

The Tyndal National Institute is supported by Research Ireland under the US-Ireland Research and Development Partnership (21/US/3755). The University of Texas at Dallas is supported by the National Science Foundation through the US-Ireland R&D Partnership Program, award number ECCS 2154535. Authors Vasily Lebedev is supported by research Ireland grant: SFI/21/US/3785 and Lewys Jones is supported by Research Ireland grants: 12/RC/2278_P2 and URF/RI/191637.

ACKNOWLEDGMENT

The authors gratefully acknowledge Brendan Sheehan and Davinder Singh from Tyndall National Institute for FIB preparation of the TEM samples. TEM microscopy characterization and analysis were performed at the CRANN Advanced Microscopy Laboratory, a Research Ireland supported imaging and analysis center.

**Figures:**

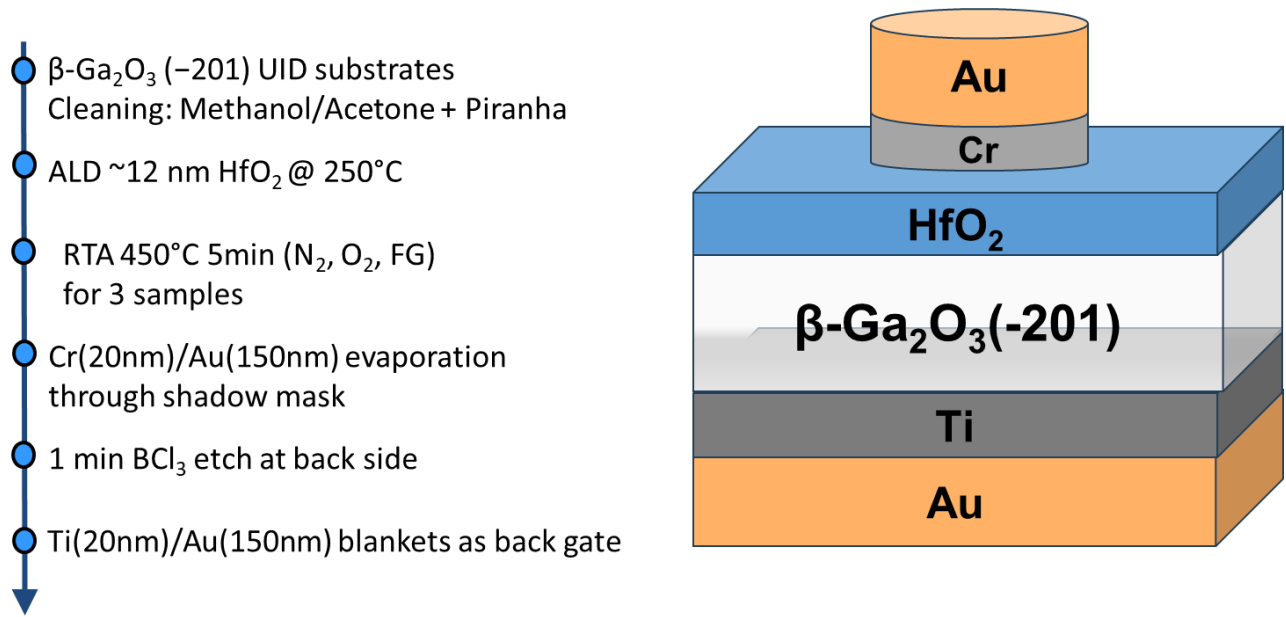

**Figure 1. Schematic diagram of the Au/Cr/HfO$_2$/Ga$_2$O$_3$ MOS structure and the fabrication process steps.**

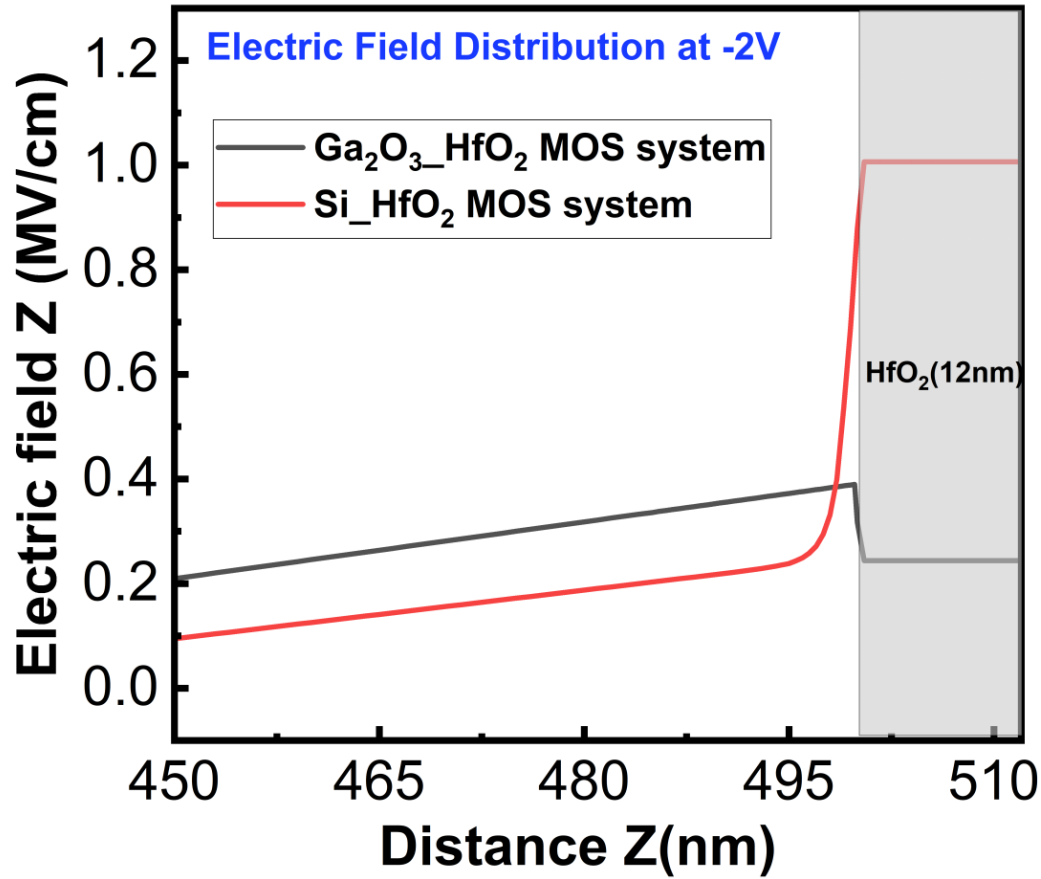

**Figure 2.** Electric field distribution of the Cr/HfO$_2$/Si and Cr/HfO$_2$/Ga$_2$O$_3$ MOS systems plotted in depletion/inversion region at -2V. The simulation is performed without considering any defects (ideal mode) Z is the distance in nm, HfO$_2$ thickness is the same for both systems.

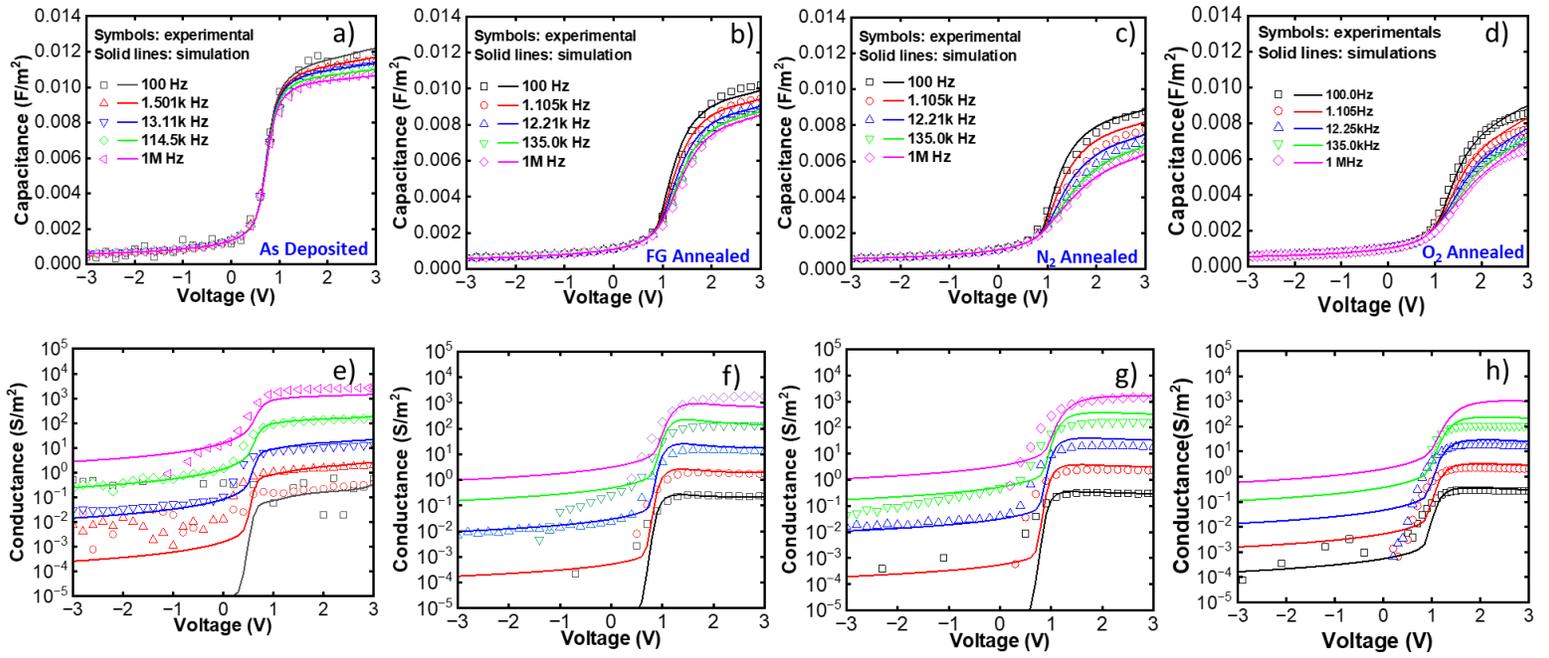

**Figure 3.** Capacitance voltage (C-V) (a-d) and conductance voltage (G-V) (e-h) curves of Au/Cr/HfO$_2$/Ga$_2$O$_3$ MOS systems measured at multiple frequencies a) As deposited b) Forming gas annealed, c) N$_2$ annealed and d) O$_2$ annealed. The experimental data is shown in symbols, while solid lines show the simulated data.

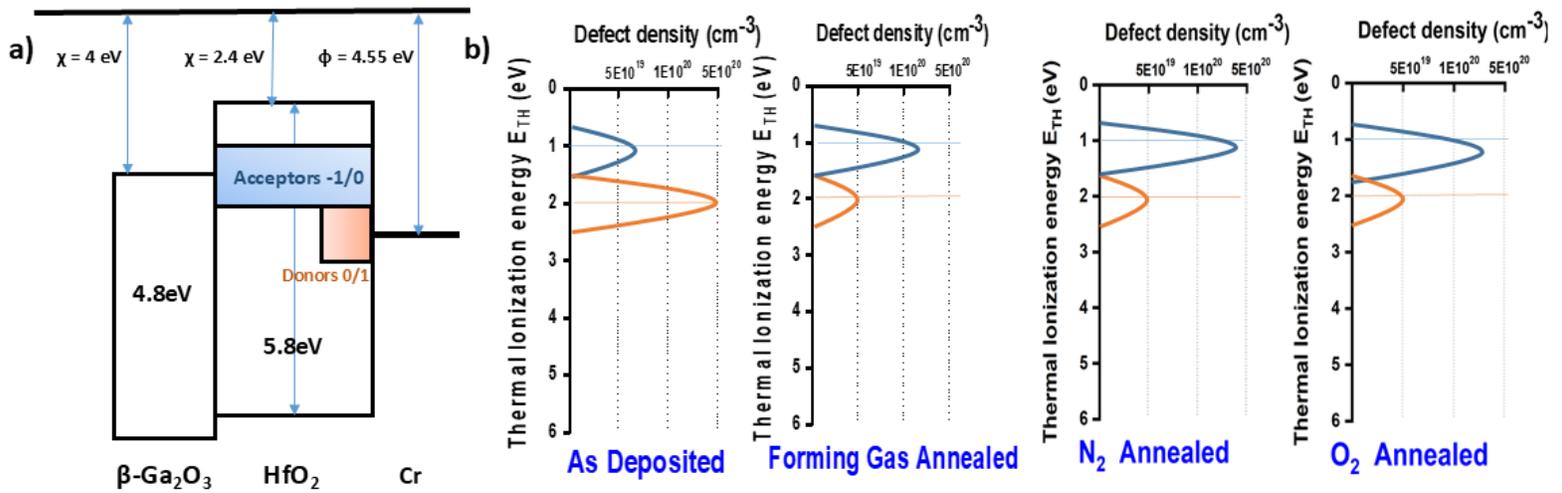

**Figure 4.** Defect distribution extracted for the Au/Cr/HfO$_2$/Ga$_2$O$_3$ MOS system. a) Band diagram of the Cr/HfO$_2$/Ga$_2$O$_3$ stack with two distinct defect distributions, the acceptor band is distributed throughout the HfO$_2$ thickness, while the donor band is confined near the HfO$_2$/Cr interface. b) Defect volume density of the acceptor and donor bands plotted as function of thermal ionization energy for as deposited, FG annealed, N$_2$ and O$_2$ annealed samples.

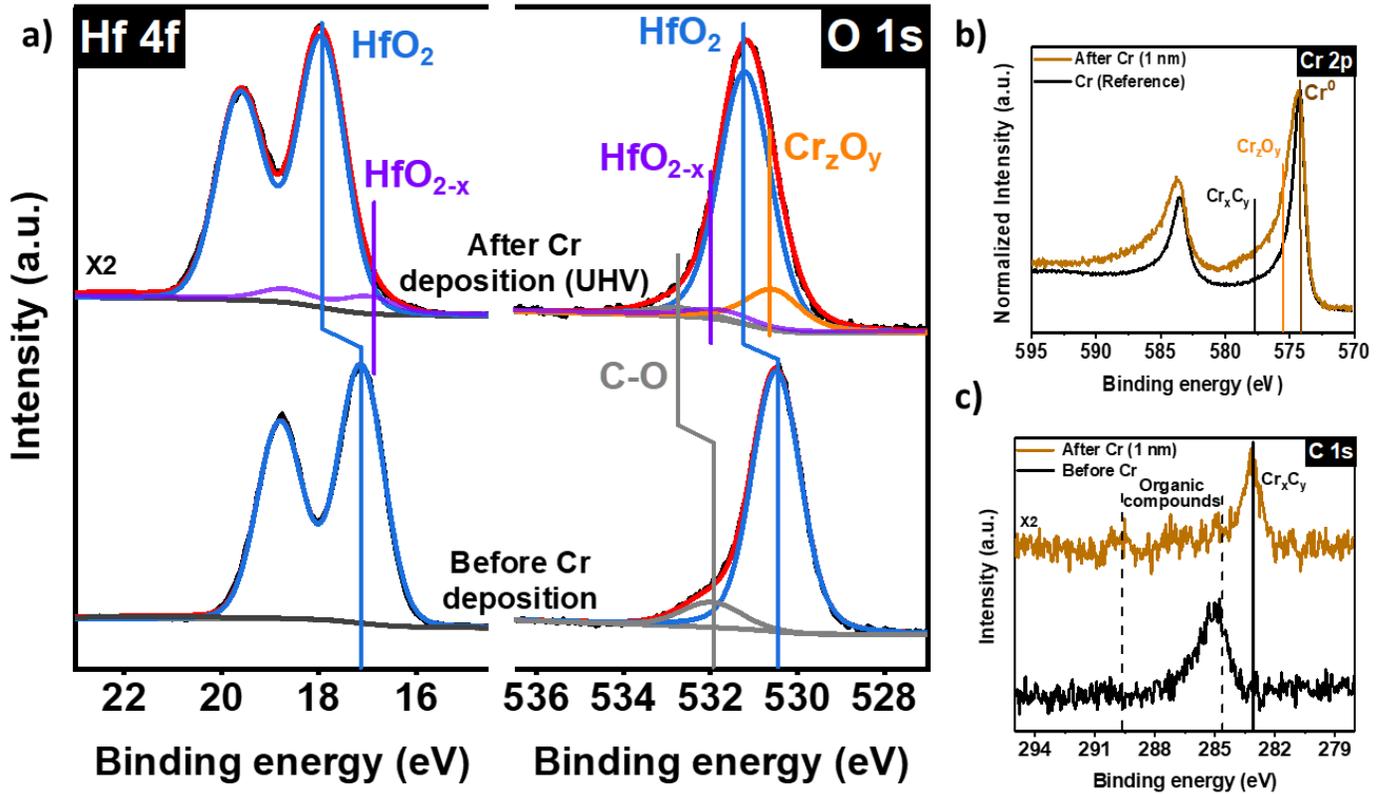

**Figure 5:** In situ XPS of the Cr/HfO₂ interface on β-Ga₂O₃ ($\bar{2}$01) bulk substrate with 12 nm of ALD- grown HfO₂ (a) Hf *4f* O *1s*, (b) Cr *2p* and (c) C *1s* interface chemistry before and after UHV Cr gate metal deposition. Intensities are multiplied 2x after the Cr deposition as the overlayer obscures the underlying elemental states.

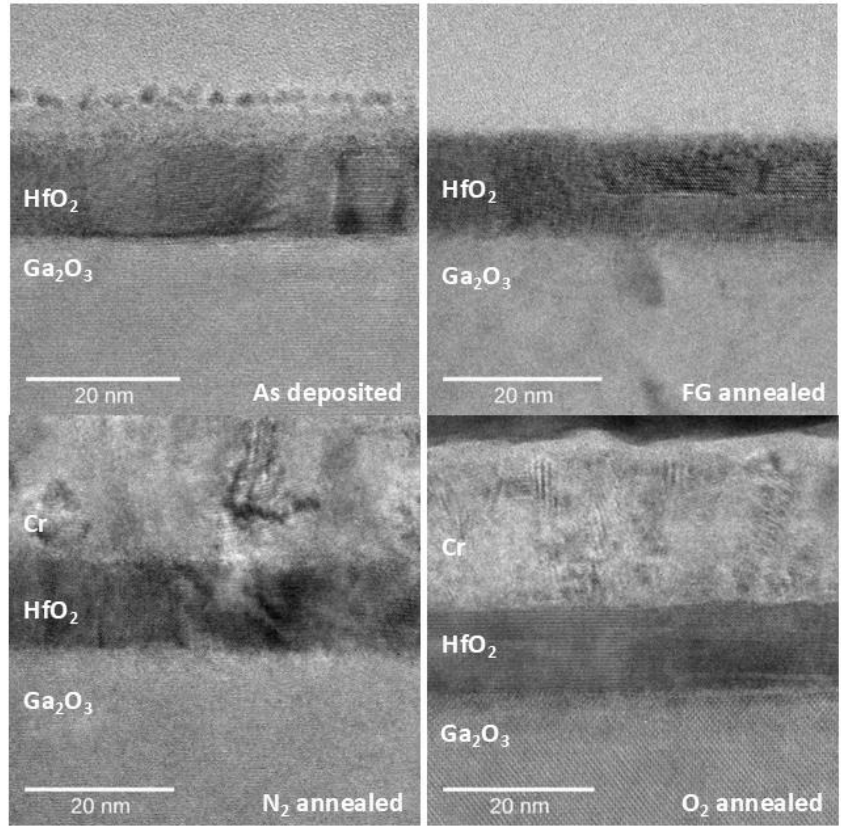

**Figure 6.** High-resolution cross-sectional TEM images of the Au/Cr/HfO$_2$(12nm)/Ga$_2$O$_3$ MOS for as deposited, FG annealed, N$_2$ and O$_2$ annealed sample. The observed HfO$_2$ layer thickness for all four samples is ~11.5 ± 0.5 nm. Note the absence of Cr above HfO$_2$ for the as deposited and FG annealed samples because the TEM images were obtained outside the capacitor area.

**Supporting Information**

# Defects in the β-Ga$_2$O$_3$($\bar{2}$01)/HfO$_2$ MOS system and the effect of thermal treatments

Khushabu. S. Agrawal[1], Paolo LaTorraca[1], Jonas Valentijn[1], Roberta Hawkins[2,‡], Adam A. Gruszecki[2], Joy Roy[2], Vasily Lebedev[3], Lewys Jones[3], Robert M. Wallace[2], Chadwin D. Young[2], Paul K. Hurley[1] and Karim Cherkaoui[1,*]

[1]Tyndall National Institute, University College Cork, Lee Maltings, Prospect Row, Cork, Ireland

[2]Department of Materials Science and Engineering, The University of Texas at Dallas, Richardson, Texas, 75080, United States

[4]Advanced Microscopy Laboratory (AML), School of Physics, CRANN & AMBER Trinity College Dublin, the University of Dublin, Dublin 2, Ireland

[‡]Now at Qorvo, Inc., Richardson, Texas, 75080, United States

**XPS Analysis:**

For the X-ray photoelectron spectroscopy (XPS) analysis, β-$Ga_2O_3$ ($\bar{2}01$) bulk sample was standard cleaned using methanol and acetone followed by a 5 min dip in piranha solution and 5 min 10:1 HF dip, finishing with a thorough wash in deionized water. Within 5 minutes of exposure time, the sample was loaded using a stainless-steel sample holder plate into a load lock chamber attached to a cluster system[1]. The in-situ XPS was performed in as-loaded conditions

After the initial scan, the crystal was transferred through the transfer tube (base pressure, $P_b \approx 5 \times 10^{-11}$ mbar) under ultra-high vacuum (UHV) to an atomic layer deposition (ALD) chamber for the deposition of 12 nm $HfO_2$. Thereafter, the sample was transferred back to the analytical chamber for studying the $HfO_2$ surface composition. Later, the sample was transferred to a physical vapor deposition chamber ($P_b \approx 3 \times 10^{-11}$ mbar) where an electron beam deposition technique was used to deposit gate metal. Approximately 1 nm Cr layer was deposited at room temperature at a deposition rate of 0.01 nm/sec. Final XPS scans were done after UHV metallization and interface chemistries were analyzed.

**Figure S1:**

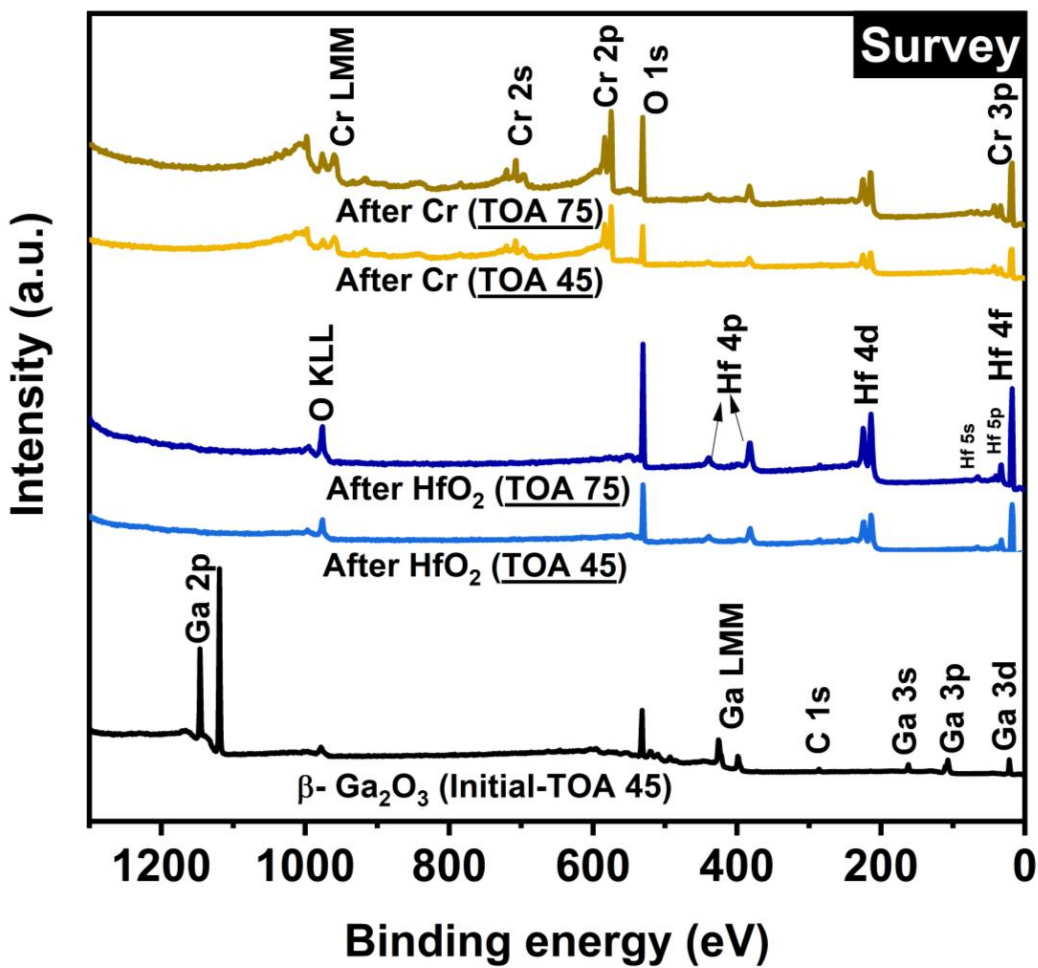

Figure S1: A survey scan including the initial scan of β-Ga$_2$O$_3$ ($\bar{2}$01) substrate as loaded, and after subsequent ALD HfO$_2$ growth and UHV metallization, performed in situ at take-off angles 45º and 75º. The thicknesses of the HfO$_2$ and Cr are 12 nm and 1 nm respectively.

**Figure S2:**

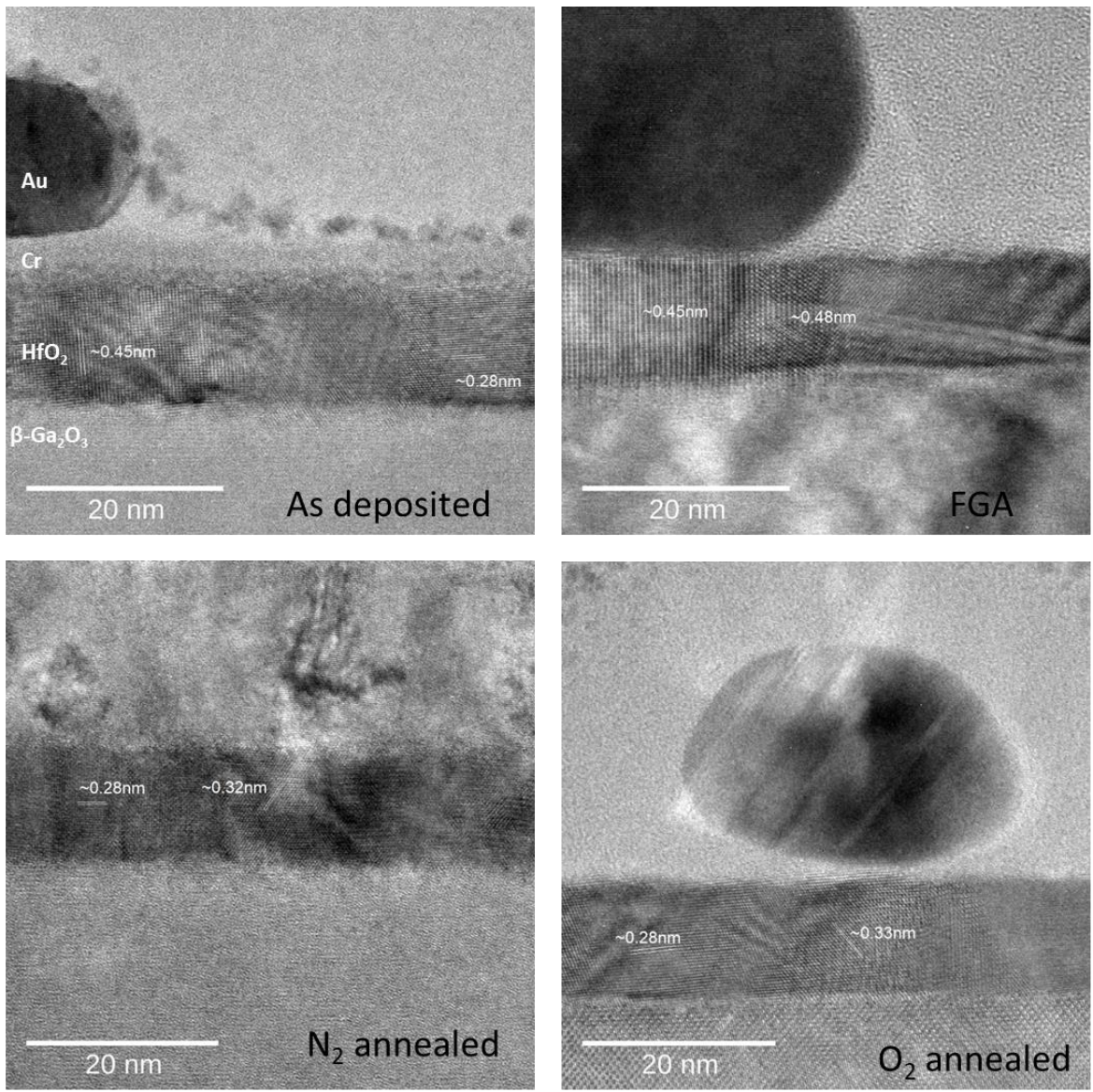

Figure S2: Cross sectional HR-TEM images of different samples showing the polycrystalline $HfO_2$. The as deposited and Forming gas annealing (FGA) lamellas were taken at the edge of the capacitors showing the edge of the Au metallisation.

**Figure S3:**

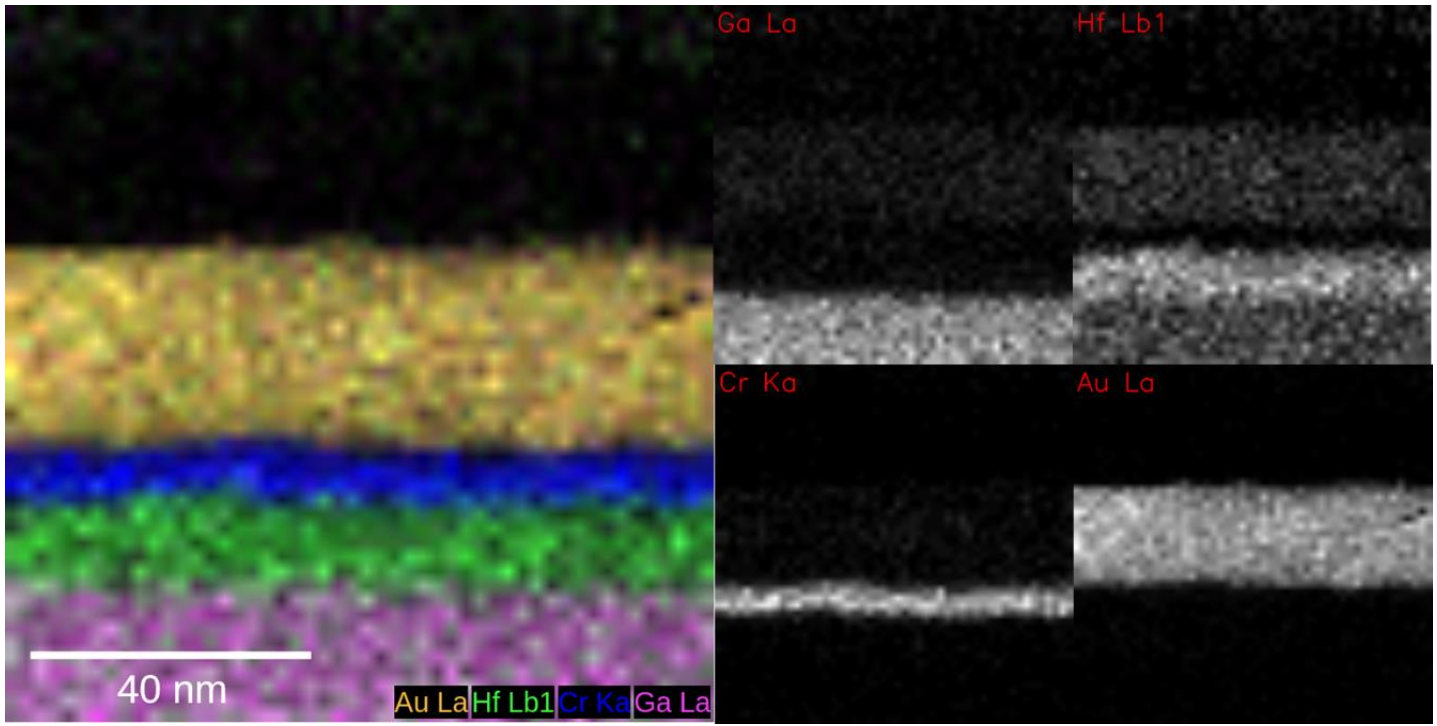

Figure S3: Elemental analysis of $Ga_2O_3$/$HfO_2$/Cr/Au MOS layers using X-ray energy dispersive spectroscopy (EDS) for the control sample. No significant diffusion of Ga into the $HfO_2$ layer is observed in all the acquired maps.

**Figure S4:**

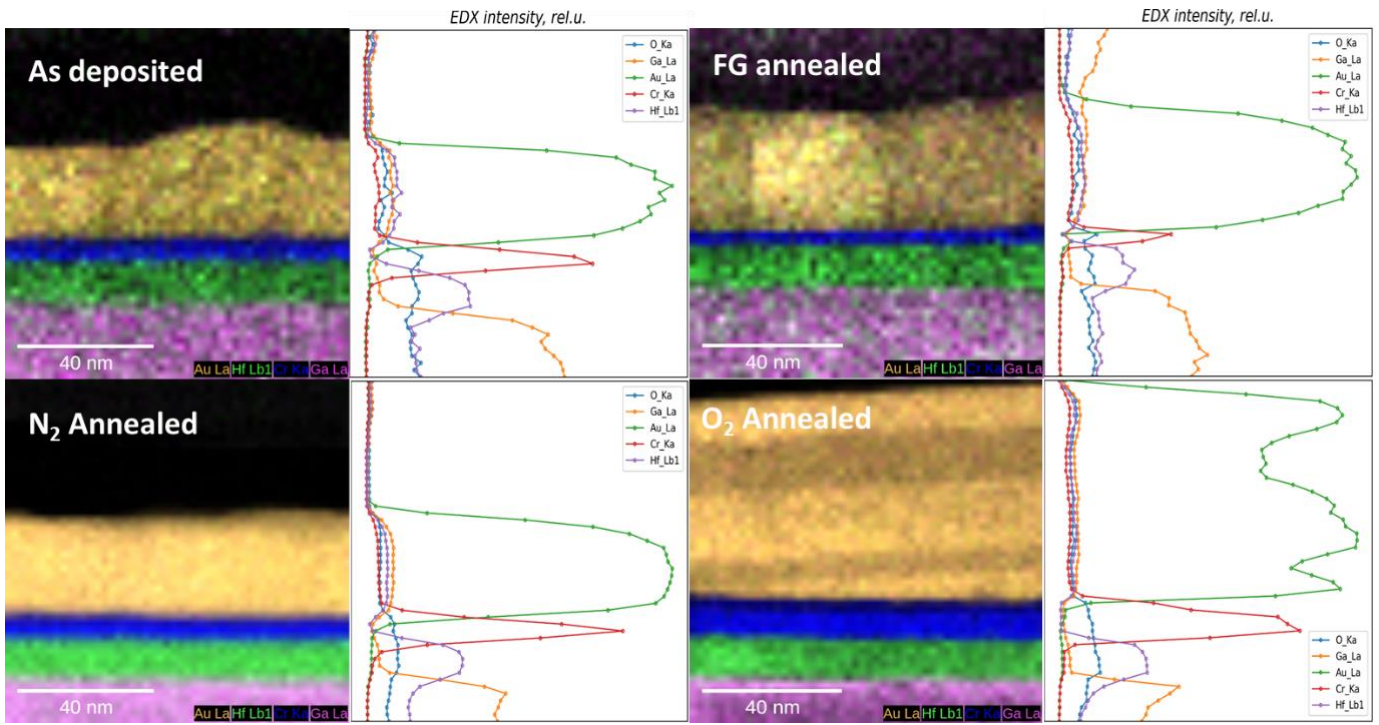

Figure S4: EDS integrated line profiles showing the relative intensity of O, Ga, Au, Cr and Hf signals. The observed O K-edge peak intensity in the Cr layer is affected by the correlation between Cr L and O K energies for all four samples.

**Figure S5:**

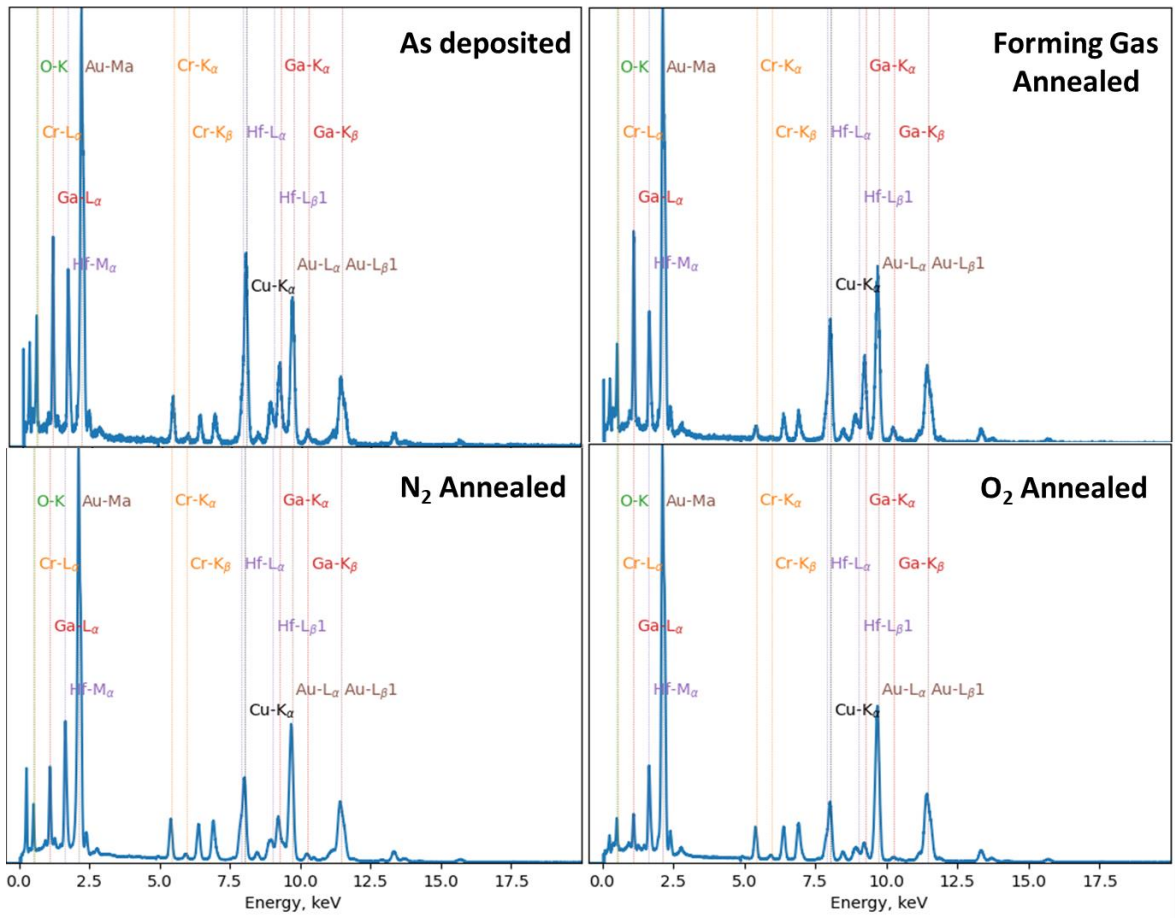

Figure S5: Cumulative EDS spectrum for all 4 samples